\documentclass{article}
\usepackage[dvips]{graphicx}
\usepackage{amsmath}

\begin{document}

\begin{center}
{\textit {\textbf{T.G.Elizarova, I.S.Kalachinskaya, Yu.V.Sheretov
 }}}

 \vspace{1cm}

 {\Large {\textbf{ Separating Flow Behind a Back-Step. Part I.
Quasi-Hydrodynamic Equations and Computation of a Laminar Flow}}}
\end{center}

We demonstrate the results of the numerical modelling of a plane
two-dimensional viscous incompressible flow in a channel with a
back-step. As a mathematical model we take equations for a
incompressible flow based on the quasi-hydrodynamic (QHD)
equations. We present a phenomenological derivation of the QHD
equations and show a relation of these equations to the
Navier-Stokes system. We test the proposed numerical algorithm by
computing of certain laminar flows.


\section{Introduction}

The paper is devoted to the numerical modelling of the viscous
incompressible flow behind the back-step in the channel with a
sudden broadening. The quasi-hydrodynamic (QHD) system of
equations is used as a mathematical model. \cite{Sher97},
\cite{Sher2000}.

In the second part of our paper we discuss the derivation of the
QHD equations and their relation to the Navier-Stokes equations.
The QHD equations broaden the possibilities of the classical
Navier-Stokes model in description of the viscous compressible gas
flows. When the Navier-Stokes equations are applicable, the
additional dissipation of the QHD equations makes little influence
on the solutions, but provides the stability of numerical
computations. In certain cases of weakly rarified flows the QHD
equations give the solution that describes experimental data
better than the Navier-Stokes model \cite{Microcan}.

Probation of the QHD equations for computing the incompressible
liquid flows and for the problems of thermal and thermocapillar
convection was carried out in \cite{JVM2001}--\cite{JVM2003}. In
particular, it was shown that the QHD equations are effective for
modelling of nonstationary flows \cite{JVM98}.

The size of the separation zone behind the step is a sensitive
characteristic feature of laminar flows, which strongly depends on
the flow velocity and on the geometry of the region considered in
the problem. Analytical expression for the dependence of the
separation zone's length on the Reynolds number and the relative
height of the step for two-dimensional flows is given, for
example, in \cite{Sparrow}. The length of the separation zone
grows almost linearly with the increasing of the Reynolds number.
Laminar flows behind the back step are well simulated numerically
and the results of two-dimensional computations of different
authors are in a good agreement with experimental data
\cite{Sparrow}--\cite{Hack}. It allows us to use this problem as a
test for probation of new numerical algorithms.

In this paper we present the computations of the laminar flows in comparison
with the previously published results. Data obtained from literature is used
for
evaluating the robustness and the accuracy of the numerical algorithm for
computing the flow behind the step, which is based on the QHD equations. In
the
continuation of this paper (Part II) the method proposed here is applied to
the
numerical investigation of turbulent flows behind the step.

\section{Mathematical model}

In this section we describe the physical principles that form the
basis for the phenomenological derivation of the new
quasi-hydrodynamic (QHD)system of equations. Using time-space
averaging for introducing the principal hydrodynamic values --
density, velocity and temperature -- is the essential and
fundamental feature that distinguishes our method from the
Navier-Stokes theory, where hydrodynamic values are introduced
based on space averaging.

\subsection{Integral conservation laws}
Let us consider an inertial Cartesian coordinate system $(x_1,
x_2, x_3)$ in the Euclidian space $R_{\vec{x}}^3$ . Let
$(\vec{e}_1, \vec{e}_2, \vec{e}_3)$ be the corresponding
orthonormal basis of unit vectors and let us denote time as $t$.
We shall use the following standard notation for the variables
describing the viscous compressible thermoconducting flow:
$\rho=\rho(\vec{x},t)$ -- density, $\vec{u}=\vec{u}(\vec{x},t)$ --
velocity, $p=p(\vec{x},t)$ -- pressure,
$\varepsilon=\varepsilon(\vec{x},t)$ -- specific internal energy,
$T=T(\vec{x},t)$ -- temperature, $s=s(\vec{x},t)$ -- specific
entropy.

Suppose that the medium is two-parametric, that is, only two out
of five thermodynamical parameters $\rho$, $p$, $\varepsilon$,
$T$, $s$ are independent, and we are given by the state equations
\begin{equation}\label{Cond1}
p=p(\rho,T),\quad \varepsilon=\varepsilon(\rho,T),\quad
s=s(\rho,T).
\end{equation}
Let $\vec{F}=\vec{F}(\vec{x},t)$ be the mass density of external forces. For
example, in case of the liquid in the gravitational field of the Earth it
will
be $\vec{F}=\vec{g}$, where $\vec{g}$ is the gravity acceleration.

Our first postulate is the law of conservation of mass in the following form:
\begin{equation}\label{Mass1}
\frac{\partial\rho}{\partial t}+\it{div }\vec{j}_m = 0.
\end{equation}
We suppose that the mass flux density vector $\vec{j}_m=\vec{j}_m(\vec{x},t)$
is
defined in every point $\vec{x}$ of the flow in every moment of time $t$. In
the
region occupied by the flow we take an arbitrary moving material volume
$V=V(t)$
with the smooth surface $\Sigma=\Sigma(t)$, oriented with the field of
external
normal unit vectors $\vec{n}$. We also suppose that the volume $V(t)$
originates
from the volume $V_0=V(t_0)$, where $t_0$ is the initial moment of time, by
continuous deformation, caused by the motion of particles $V_0$ along the
trajectories, determined by the vector field $\vec{j}_m/\rho$. Using the
well-known \cite{Loy} Euler--Liouville identity
\begin{equation}\label{Liuv}
\frac{d}{dt}\int_{V}\varphi dV = \int_{V}[D\varphi+ \varphi\it{div
}(\vec{j}_m/\rho)]dV,
\end{equation}
where $\varphi=\varphi(\vec{x},t)$ is a certain continuously
differentiable scalar or vector field, $dV$ is a volume element in
$R_{\vec{x}}^3$ and $D=\partial /\partial
t+(\vec{j}_m/\rho)\cdot\vec{\nabla}$ is the differential operator,
we present the law of conservation of mass (\ref{Mass1}) in the
integral form:
\begin{equation}\label{Mass2}
\frac{d}{dt}\int_V\rho dV = 0.
\end{equation}

The second postulate is the law of conservation of momentum
\begin{equation}\label{Mom1}
\frac{d}{dt}\int_V (\rho\vec{u}) dV = \int_V\rho\vec{F} dV +
\int\!\!\int_{\Sigma} (\vec{n}\cdot P) d\Sigma,
\end{equation}
where $d\Sigma$ is the element of the surface $\Sigma$ in the
vicinity of the unit vector $\vec{n}$. The rate of variation of
the momentum in the volume $V$ equals to the sum of all forces
applied to it. The first integral in the right hand side of
(\ref{Mom1}) is a volume force caused by the external field; the
second  stands for the forces, caused by pressure and internal
viscous friction, that are applied to the surface $\Sigma$. The
variable $P=P(\vec{x},t)$ is called the tensor of internal
tensions. The symbol $(\vec{n}\cdot P)$ means the contraction (dot
product) of the vector $\vec{n}$ and the second rank tensor $P$
with respect to the first index of the tensor. Respectively,
$(P\cdot\vec{n})$ means that the contraction of $P$ and $\vec{n}$
is done with respect to the second index of $P$. If the tensor $P$
is symmetric, then $(\vec{n}\cdot P) = (P\cdot\vec{n})$.

The third postulate is the law of conservation of the total energy
\begin{equation}\label{Energy1}
\frac{d}{dt}\int_V \rho\Bigl(\frac{\vec{u}^2}{2}+\varepsilon\Bigr)
dV = \int_V (\vec{j}_m\cdot\vec{F}) dV +
\int\!\!\int_{\Sigma}(\vec{A}\cdot\vec{n}) d\Sigma -
\int\!\!\int_{\Sigma}(\vec{q}\cdot\vec{n}) d\Sigma.
\end{equation}
Here the first integral in the right hand side of (\ref{Energy1})
equals to the capacity of the external volume forces that are
applied to the volume $V$; the second is understood as the
capacity of the surface forces of the pressure and the internal
viscous stress. The last term in (\ref{Energy1}) describes the
influx of energy in a single unit of time through the surface
$\Sigma$ due to the processes of the heat transfer. Actual
expressions for the vector fields $\vec{A}=\vec{A}(\vec{x},t)$ and
$\vec{q}=\vec{q}(\vec{x},t)$ will be given below.

The fourth postulate expresses the law of conservation of the moment of
momentum:
\begin{equation}\label{AngMom1}
\frac{d}{dt}\int_V [\vec{x}\times(\rho\vec{u})] dV = \int_V
[\vec{x}\times\rho\vec{F}] dV+
\int\!\!\int_{\Sigma}[\vec{x}\times(\vec{n}\cdot P)] d\Sigma.
\end{equation}
It is presented in its classical form. Internal moments and the distributed
mass
and surface pairs are not taken into consideration. The symbol $\times$
denotes
the cross product of two vectors.

Our fifth postulate is the second law of thermodynamics. It looks as follows:
\begin{equation}\label{Entropy1}
\frac{d}{dt}\int_V (\rho s) dV = -
\int\!\!\int_{\Sigma}\frac{(\vec{q}\cdot\vec{n})}{T} d\Sigma +
\int_V X dV.
\end{equation}
The surface integral in the right hand side (\ref{Entropy1}) defines the rate
of
the variation of entropy in the volume $V$ due to the thermal flux. It may be
both positive or negative. The last integral is always non-negative: it gives
the production of entropy due to the internal irreversible processes.

\subsection{Transfer to differential equations}

Just like in the case of the Navier--Stokes system \cite{Loy}, for
the transfer from the integral relations
(\ref{Mass2})--(\ref{Entropy1}) to the corresponding differential
ones we use the Liouville formula (\ref{Liuv}) for differentiating
the integral over the moving material volume. Doing it, we shall
suppose that all the principal macroscopic parameters of the
medium are  sufficiently smooth functions of time and spatial
coordinates. Taking in consideration that the volume $V$ is
arbitrary, we obtain differential equations for the balances of
the mass
\begin{equation}\label{Mass3}
\frac{\partial\rho}{\partial t}+\it{div }\vec{j}_m=0,
\end{equation}
of the momentum
\begin{equation}\label{Mom2}
\frac{\partial (\rho\vec{u})}{\partial t}+ \it{div
}(\vec{j}_m\otimes\vec{u})=\rho\vec{F}+\it{div }P,
\end{equation}
of the total energy
\begin{equation}\label{Energy2}
\frac{\partial}{\partial t}\Bigl[\rho\Bigl(\frac{\vec{u}^2}{2}+
\varepsilon\Bigr)\Bigr]+\it{div }\Bigl[\vec{j}_m\Bigl(
\frac{\vec{u}^2}{2}+\varepsilon\Bigr)\Bigr]=
(\vec{j}_m\cdot\vec{F})+\it{div }\vec{A}-\it{div }\vec{q},
\end{equation}
of the moment of momentum
\begin{equation}\label{AngMom2}
\frac{\partial}{\partial t}[\vec{x}\times\rho\vec{u}]+ \it{div
}(\vec{j}_m\otimes[\vec{x}\times\vec{u}]) =
[\vec{x}\times\rho\vec{F}]+\frac{\partial}{\partial x_i}
[\vec{x}\times P_{ij}{\vec{e}}_j]
\end{equation}
and of the entropy
\begin{equation}\label{Entropy2}
\frac{\partial (\rho s)}{\partial t}+ \it{div }(\vec{j}_m
s)=-\it{div }\Bigl(\frac{\vec{q}}{T}\Bigr)+X.
\end{equation}
Here $(\vec{j}_m\otimes\vec{u})$ is the second rank tensor
obtained as a direct product of the vectors $\vec{j}_m$ and
$\vec{u}$. When we take the divergence of the second rank tensor,
we carry out the contraction with respect to its first index. The
symbol $P_{ij}$ in the equation (\ref{AngMom2}) means the portrait
of the tensor $P$ in the basis $(\vec{e}_1,\vec{e}_2,\vec{e}_3).$
The summation is carried out with respect to the indexes $i$ and
$j$ that appear twice.

The system (\ref{Mass3})--(\ref{Entropy2}) is not closed.  It is
necessary to introduce the variables $\vec{j}_m$, $P$, $\vec{q}$,
$\vec{A}$, $X$ as the functions of macroscopic parameters of the
medium and their derivatives. The closure problem can be solved in
several ways.

\subsection{The classical approach to the closure problem. The Navier--Stokes
equations}

First of all let us discuss the classical approach \cite{Loy}, in which the
averaging over a certain set of physically infinitely small volumes from the
space $R_{\vec{x}}^3$ at the fixed moment of time $t$ is used for definition
of
hydrodynamic variables. In this case the mass flow density vector $\vec{j}_m$
at
the arbitrary point $(\vec{x},t)$ coincides with the average momentum of the
unit volume $\rho\vec{u}$, so the first closure relation looks as follows:
\begin{equation}\label{Jm1}
\vec{j}_m=\rho\vec{u}.
\end{equation}
After that the pressure and inner viscous friction forces are
introduced. They act instantly on the surface of the material
volume. The law of motion of the latter is chosen in the same way
as in the rigid body mechanics. In literature this assumption is
called the solidification principle. The balance equation for the
angular momentum (\ref{AngMom2}) follows from the momentum
conservation law (\ref{Mom2}) and the symmetry of the tension
tensor $P$. In the theory for Newtonian media $P=P_{NS}$ is
defined by the expression
\begin{equation}\label{Pi1}
P=\Pi_{NS}-pI,
\end{equation}
where
\begin{equation}\label{PiNS}
\Pi_{NS} = \eta
[(\vec{\nabla}\otimes\vec{u})+(\vec{\nabla}\otimes\vec{u})^{T}-
(2/3)I\it{div }\vec{u}]
\end{equation}
-- is the Navier-Stokes shear-stress tensor, $I$ -- the unit
tensor -- is the invariant of the second range. The heat flow
$\vec{q}=\vec{q}_{NS}$ is defined according to the Fourier law
\begin{equation}\label{QNS}
\vec{q}=-\it\ae\vec{\nabla}T.
\end{equation}
The hypothesis (\ref{PiNS}) and (\ref{QNS}) for the ideal
monoatomic gases with small Knudsen numbers are confirmed by the
kinetic computations. The work of the surface pressure forces and
inner viscous shear-stress  forces in a unit of time is computed
using the same formula as in the rigid body mechanics, that is:
\begin{equation}\label{ANS}
\vec{A}=(P_{NS}\cdot\vec{u}).
\end{equation}
The specific thermodynamical entropy is supposed to satisfy the Gibbs
differential identity
\begin{equation}\label{Gibbs}
Tds=d\varepsilon+pd(1/\rho).
\end{equation}
Its balance equation (\ref{Entropy2}) may be obtained as a consequence of the
mass, momentum and energy conservation laws (\ref{Mom2})--(\ref{Energy2}), if
we
choose $X=X_{NS}$ as
\begin{equation}\label{XNS}
X=\it\ae\Bigl(\frac{\vec{\nabla}T}{T}\Bigr)^{2}+
\frac{(\Pi_{NS}:\Pi_{NS})}{2\eta T},
\end{equation}
where
$(\Pi_{NS}:\Pi_{NS})=\sum_{i,j=1}^{3}(\Pi_{NS})_{ij}(\Pi_{NS})_{ij}$
-- is the double dot product of two identical tensors. Note that
the right hand side of the equality (\ref{XNS}) is non-negative.
Substitution of the expressions (\ref{Jm1})--(\ref{ANS}) into
equations (\ref{Mass3}) -- (\ref{Mom2}) gives us the classical
Navier-Stokes system. The dependensies $\eta=\eta(\rho,T)$ and
$\ae=\ae(\rho,T)$ may be either found experimentally or derived
from the kinetic theory of gases.

\subsection{The non-traditional approach to the closure problem. The
quasi-hydrodynamic system}

Another way of solving the problem of closing the system
(\ref{Mass3})--(\ref{Entropy2}) was proposed by Yu.V.Sheretov in
\cite{Sher97}, \cite{Sher2000}. To define hydrodynamic variables
he used not the spatial, but the time-spatial averaging over a
certain set of physically infinitely small four-dimensional
volumes in the space $R_{\vec{x},t}^4$. He has proved that in the
case of such time-spatial averaging the mass flow density vector
$\vec{j}_m$, generally speaking, doesn't coincide with the average
momentum of the unit volume $\rho\vec{u}$. Detailed analysis of
different possibilities for choosing the variables $\vec{j}_m$,
$P$, $\vec{q}$, $\vec{A}$ Õ $X$ gave the following result:
\begin{equation}\label{JmQHD}
\vec{j}_m=\rho(\vec{u}-\vec{w}),
\end{equation}
\begin{equation}\label{PiQHD}
P=-pI+\Pi_{NS}+\rho\vec{u}\otimes\vec{w},
\end{equation}
\begin{equation}\label{QQHD}
\vec{q}=-\it\ae\vec{\nabla}T,
\end{equation}
\begin{equation}\label{AQHD}
\vec{A}=(\Pi_{NS}\cdot\vec{u})+
\rho\vec{u}(\vec{w}\cdot\vec{u})-p(\vec{u}-\vec{w}),
\end{equation}
\begin{equation}\label{XQHD}
X=\it\ae\Bigl(\frac{\vec{\nabla}T}{T}\Bigr)^{2}+
\frac{(\Pi_{NS}:\Pi_{NS})}{2\eta T}+\frac{\rho\vec{w}^2}{\tau T},
\end{equation}
where
\begin{equation}\label{WQHD}
\vec{w}=\frac{\tau}{\rho}[\rho(\vec{u}\cdot\vec{\nabla})\vec{u}+
\vec{\nabla}p-\rho\vec{F}].
\end{equation}
The parameter $\tau=\tau(\rho,T)$ describes the scale of temporal smoothing.
The
formula for computing this parameter was proposed in \cite{Sher2000}:
\begin{equation}
\label{Cond2} \tau=\frac{\gamma}{Sc}\frac{\eta}{\rho c_s^2},
\end{equation}
where $\gamma$ is the isentropic exponent, $Sc$ is the Schmidt
number (which is close to 1 for gases), $c_s$ is the sound
velocity.
  The value of $\tau$ agrees by order with the
average mean free path of the particles in gas. Computations for
moderately rarified gases confirm the correctness of this choice
of the smoothing parameter \cite{Microcan}.

Having substituted the expressions (\ref{JmQHD}), (\ref{PiQHD}) and
(\ref{AQHD})
instead of $\vec{j}_m$, $P$ and $\vec{A}$ in (\ref{Mass3})--(\ref{Energy2}),
we
obtain the quasi-hydrodynamic (QHD) system of equations:
\begin{equation}\label{Mass4}
\frac{\partial\rho}{\partial t}+\it{div }(\rho\vec{u})= \it{div
}(\rho\vec{w}),
\end{equation}
\begin{equation}\label{Mom3}
\frac{\partial(\rho\vec{u})}{\partial t}+ \it{div
}(\rho\vec{u}\otimes\vec{u})+\vec{\nabla}p=\rho\vec{F}+ \it{div
}\Pi_{NS}+ \it{div
}[(\rho\vec{w}\otimes\vec{u})+(\rho\vec{u}\otimes\vec{w})],
\end{equation}
\begin{eqnarray}\label{Energy3}
&&\frac{\partial}{\partial t}\Bigl[\rho\Bigl(\frac{\vec{u}^2}{2}+
\varepsilon\Bigr)\Bigr]+\it{div }\Bigl[\rho\vec{u}\Bigl(
\frac{\vec{u}^2}{2}+\varepsilon\Bigr)+p\vec{u}\Bigr]+ \it{div
}\vec{q}=\rho\vec{F}\cdot(\vec{u}-\vec{w})+\nonumber\\ &&+\it{div
}(\Pi_{NS}\cdot\vec{u})+ \it{div
}\Bigl[\rho\vec{w}\Bigl(\frac{\vec{u}^2}{2}+
\varepsilon\Bigr)+p\vec{w}+\rho\vec{u}(\vec{w}\cdot\vec{u})\Bigr].
\end{eqnarray}
The QHD system (\ref{Mass4})--(\ref{Energy3}) becomes closed if it is
equipped
with the state equations (\ref{Cond1}), and the coefficients $\eta$, $\it\ae$
and $\tau$ are presented as functions of macroscopic parameters of the media.
The substitution of the expressions (\ref{JmQHD}), (\ref{QQHD}) and
(\ref{XQHD})
into (\ref{Entropy2}) gives the entropy balance equation
\begin{equation}\label{Entropy3}
\frac{\partial(\rho s)}{\partial t}+\it{div }(\rho\vec{u}s)=
\it{div }(\rho\vec{w}s)+ \it{div
}\Bigl(\it\ae\frac{\vec{\nabla}T}{T}\Bigr)+
\it\ae\Bigl(\frac{\vec{\nabla}T}{T}\Bigr)^{2}+\frac{\Psi_{QHD}}{T},
\end{equation}
in which $$
\Psi_{QHD}=\frac{(\Pi_{NS}:\Pi_{NS})}{2\eta}+\frac{\rho\vec{w}^2}{\tau}
$$ is the non-negative dissipative function.

A number of theoretical results is obtained for the QHD system
(\ref{Mass4})--(\ref{Energy3}) in \cite{Sher97}, \cite{Sher2000}. In
particular,
it has been shown that the stationary QHD system in dimensionless variables
differs from the corresponding Navier-Stokes equations only in terms of the
second order of magnitude with respect to the Knudsen number. Its laminar
boundary layer approximation is the classical Prandtl system.

\subsection{Quasi-hydrodynamic system for a viscous incompressible fluid}

In many particular cases of hydrodynamic flows we may neglect the
density variation. Supposing that $\rho$ and $T$ are constant,
from the equations (\ref{Mass4}), (\ref{Mom3}) we obtain the
system
\begin{equation}\label{Mass5}
\it{div }\vec{u}=\it{div }\vec{w},
\end{equation}
\begin{equation}\label{Mom4}
\frac{\partial\vec{u}}{\partial t}+\it{div
}(\vec{u}\otimes\vec{u})+ \frac{1}{\rho}\vec{\nabla}
p=\vec{F}+\frac{1}{\rho}\it{div }\Pi_{NS}+ \it{div
}\bigl[(\vec{w}\otimes\vec{u})+(\vec{u}\otimes\vec{w})\bigr],
\end{equation}
which is closed with respect to the unknown functions - the velocity
$\vec{u}=\vec{u}(\vec{x},t)$ and the pressure  $p=p(\vec{x},t)$. Here the
vector

$\vec{w}$ is defined by formula
$$
\vec{w}=\tau\Bigl((\vec{u}\cdot\vec{\nabla})\vec{u}+
\frac{1}{\rho}\vec{\nabla} p-\vec{F}\Bigr).
$$
We shall compute the tensor $\Pi_{NS}$ using the expression
$$
\Pi_{NS} = \eta [(\vec{\nabla}\otimes\vec{u})+
(\vec{\nabla}\otimes\vec{u})^{T}].
$$
The coefficient of dynamical viscosity $\eta$ and the
characteristic time $\tau$ are considered to be given positive
constants. Taking the formal limits in (\ref{Mass5})--(\ref{Mom4})
as $\tau\to 0$, we get the classical Navier- Stokes equations that
describe the viscous non-compressible flows. The system
(\ref{Mass5})--(\ref{Mom4}) is dissipative and possesses several
explicit and physically reasonable solutions
\cite{Sher97}--\cite{JVM2001}. The particular case of plane or
spatial axially symmetric isothermal flows without external forces
$\vec F =0$ gives
\begin{eqnarray}\label{Mass6}
\frac{\partial u_x}{\partial x}+
\frac{1}{y^k}\frac{\partial (y^k u_y)}{\partial y}=
\frac{\partial w_x}{\partial x}+
\frac{1}{y^k}\frac{\partial (y^k w_y)}{\partial y},
\end{eqnarray}
\begin{eqnarray}\label{Mom5}
&&\frac{\partial u_x}{\partial t}+\frac{\partial (u_x^2)}{\partial
x}+ \frac{1}{y^k}\frac{\partial (y^k u_y u_x)}{\partial y}+
\frac{1}{\rho}\frac{\partial p}{\partial x}=\nonumber\\
&&=2\frac{\partial}{\partial x}\Bigl(\nu\frac{\partial
u_x}{\partial x}\Bigr)+ \frac{1}{y^k}\frac{\partial}{\partial y}
\Bigl[y^k\nu\Bigl(\frac{\partial u_x}{\partial y}+
\frac{\partial u_y}{\partial x}\Bigr)\Bigr]+\nonumber\\
&&+2\frac{\partial (u_x w_x)}{\partial x}+
\frac{1}{y^k}\frac{\partial(y^k u_y w_x)}{\partial y}+
\frac{1}{y^k}\frac{\partial (y^k u_x w_y)}{\partial y},
\end{eqnarray}
\begin{eqnarray}\label{Mom6}
&&\frac{\partial u_y}{\partial t}+ \frac{\partial (u_x
u_y)}{\partial x}+ \frac{1}{y^k}\frac{\partial (y^k
u_y^2)}{\partial y}+
\frac{1}{\rho}\frac{\partial p}{\partial y}=\nonumber\\
&&=\frac{\partial}{\partial x}\Bigl[\nu\Bigl(\frac{\partial
u_x}{\partial y} +\frac{\partial u_y}{\partial x}\Bigr)\Bigr]+
\frac{2}{y^k}\frac{\partial}{\partial y}
\Bigl(y^k\nu\frac{\partial u_y}{\partial y}\Bigr)-
2k\nu\frac{u_y}{y^2}+\nonumber\\
&&+\frac{\partial (u_x w_y)}{\partial x}+ \frac{\partial (u_y
w_x)}{\partial x}+ \frac{2}{y^k}\frac{\partial (y^k u_y
w_y)}{\partial y},
\end{eqnarray}
where $$ w_x=\tau\Bigl(u_x\frac{\partial u_x}{\partial x}+
u_y\frac{\partial u_x}{\partial y}+ \frac{1}{\rho}\frac{\partial
p}{\partial x}\Bigr),\quad w_y=\tau\Bigl(u_x\frac{\partial
u_y}{\partial x}+ u_y\frac{\partial u_y}{\partial y}+
\frac{1}{\rho}\frac{\partial p}{\partial y}\Bigr). $$ Here $\nu =
\eta/\rho$ is the coefficient of kinematic viscosity, the
parameter $k$ equals to zero in the plane case and equals to one
in the axially symmetric one. The unknown variables are the
components of the velocity $u_y=u_y(x,y,t)$, $u_x=u_x(x,y,t)$ with
respect to the ortonornal local basis $(\vec{e}_x,\vec{e}_y)$ and
the pressure $p=p(x,y,t)$. The pressure field is defined using the
already found fields of velocity and temperature by solving the
Poisson equation:
\begin{eqnarray}\label{Poisson}
&&\frac{1}{\rho}\Bigl[\frac{\partial^2 p}{\partial x^2}+
\frac{1}{y^k}\frac{\partial}{\partial y}
\Bigl(y^k \frac{\partial p}{\partial y}\Bigr)\Bigr]=
\frac{1}{\tau}\Bigl[\frac{\partial u_x}{\partial x}+
\frac{1}{y^k}\frac{\partial (y^k u_y)}{\partial y}\Bigr]-\nonumber\\
&&-\frac{\partial}{\partial x}\Bigl(u_x\frac{\partial u_x}{\partial x}+
u_y\frac{\partial u_x}{\partial y}\Bigr)
-\frac{1}{y^k}\frac{\partial}{\partial y}
\Bigl[y^k\Bigl (u_x\frac{\partial u_y}{\partial x}+
u_y\frac{\partial u_y}{\partial y}\Bigr)\Bigr],
\end{eqnarray}
This equation is the equivalent representation of (\ref{Mass6}) when
$\tau=const$.

\section{Problem statement and computational algorithm}

\begin{figure}[hb!]
\begin{center}
\includegraphics[width=.5\textwidth]{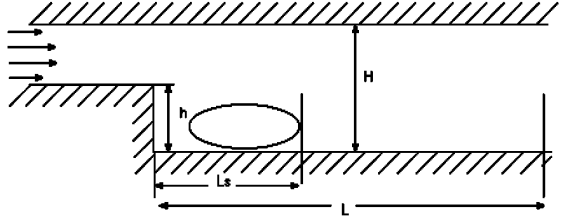}
\caption{Scheme of the computational domain} \label{Fig1}
\end{center}
\end{figure}

Let us consider a plane two-dimensional incompressible flow in the
channel of height $H$ and of length $L$ with small Mach numbers.
The channel has a narrowing at the entrance section. The size of
the narrowing is determined by the height of the step $h$. The
scheme of the computational domain and the forming flow are
demonstrated in Fig.\ref{Fig1}.

We use the QHD system (\ref{Mass6})--(\ref{Mom6}) with $k=0$ as
the  mathematical model. We transform this system into the
dimensionless form, applying the relations $$ x={\tilde x}H,\quad
y={\tilde y}H, \quad u_x={\tilde u_x}U_0,\quad u_y={\tilde
u_y}U_0,\quad p={\tilde p}\rho U_0^2,\quad t=({\tilde
t}H)/U_0,\quad Re=(U_0 H)/\nu,\quad $$ where $$ U_0 =
\frac{1}{H-h}\int_h^H u_0(y)dy $$ is the flow velocity in the
channel, averaged over the section, and $u_0(y)$ is the given
velocity profile at the entrance section. We equip this
dimensionless system with boundary conditions

$\bullet$ the solid lower wall
$$
y=0,\quad 0<x<L/H,\quad u_x = u_y =0,\quad \frac{\partial p}{\partial y}=0;
$$

$\bullet$ the solid upper wall
$$
y=1,\quad 0<x<L/H,\quad u_x = u_y =0,\quad \frac{\partial p}{\partial y}=0;
$$

$\bullet$ the solid left wall
$$
x=0,\quad 0<y<h/H,\quad u_x = u_y =0, \quad \frac{\partial p}{\partial x}=0;
$$

$\bullet$ the inflow region at the left boundary
$$
x=0,\quad h/H<y<1,\quad u_x = u_0 (y),\quad u_y =0,\quad
\frac{\partial p}{\partial x}=const;
$$

$\bullet$ the right boundary $$ x=L/H,\quad 0<y<1,\quad \frac{\partial
u_x}{\partial x}=\frac{\partial u_y}{\partial x}=0,\quad p=0. $$

Pressure boundary condition at the solid walls follows from the
non-flow conditions for the velocity components and from the
impermeability condition for the mass flow $\vec j_m$
(\ref{JmQHD}). The pressure gradient at the channel entrance may
be taken arbitrary. For example, it is possible to compute its
values in the following way: we set the velocity profile at the
channel entrance as the Poiseuille parabola \cite{Loy},
\cite{Lan}:

\begin{equation}\label{Puaz}
u_0(y)= \frac{Re}{2}\frac{\partial p}{\partial x}(1-y)(h/H-y).
\end{equation}
The mass flow rate at the entrance section is computed according
to the following formula
\begin{equation}\label{J}
J=\int_{h/H}^1 [u_x(0,y)-w_x(0,y)] dy =
-\frac{Re}{12}(1-h/H)^3\frac{\partial p}{\partial x}-
\tau(1-h/H)\frac{\partial p}{\partial x}.
\end{equation}
From (\ref{J}) we find
\begin{equation}\label{Gradp}
\frac{\partial p}{\partial x}=-\frac{12 J}{Re (1-h/H)^3}
\Bigl[1+\frac{12\tau}{Re (1-h/H)^2}\Bigr]^{-1}.
\end{equation}

We chose the initial condition: $u_x=u_y=0$. The pressure gradient
at the initial moment was supposed to be constant all over the
flow field.

The dimensionless smoothing parameter $\tau$ for laminar flows
(\ref{Cond2}) was taken equal to
\begin{equation}\label{tau1}
\tau = \frac{\gamma}{Sc}\frac{Ma}{Re_s} +\tau_0, \mbox{  where  } Ma
= \frac{U_0}{c_s} \mbox{, }  Re_s=\frac{c_s H}{\nu}
\end{equation}
- are the Mach number and the Reynolds number, derived from the
speed of sound. For example, the air at normal temperature yields
$c_s = 3.4 \cdot 10^4$ ßË/ß, $\nu = 0.15$ßË$^2$/c, $H = 10$ßË,
$Re_s = 2\cdot 10^6$. For laminar flows we have $Ma<<1$. So for
real flows the smoothing parameter proves to be small. We added to
it the value $\tau_0$ in order to compensate the difference
scheme's antidiffusion and to provide stable computing. The value
of $\tau_0$ was chosen proportional to $1/Re$.

 The QHD equations are solved numerically using the
algorithm, similar to the one described in
\cite{JVM2001}--\cite{JVM98}, - the explicit finite- difference
scheme with second order of accuracy with respect to all spatial
variables. Velocity and pressure values are defined in the same
 grid points. At each time step, the pressure field is calculated
 by using the velocity field, as a solution of Poisson equation
 (\ref{Poisson}), which is also approximated with the second order space
 accuracy. The Poisson equation is solved by the preconditioned
 generalized conjugate gradient method.

  To present the numerical results, let us also introduce
the stream-function, which is related to the solenoidal field
${\vec u} - {\vec w}$. These relations \cite{Loy} look as follows:
\begin{equation}\label{streamf}
u_y - w_y =-\frac{1}{y^k}\frac{\partial \psi}{\partial x}, \quad
u_x - w_x =\frac{1}{y^k}\frac{\partial \psi}{\partial y}.
\end{equation}
The boundary conditions for the stream-function are defined as
follows - At the lower boundary of the computational domain and at
the left wall we use the normalization $\psi = 0$, because there
we have the impermeable boundary conditions. At the upper boundary
 the stream-function equals to the mass flow rate of the liquid.

\section{Numerical modeling of laminar flows}

For proper verification of the numerical method for back-step flow
the problem described above has been solved with $Re$=100, 200,
300, 400; $h/H=1/2$. (From here on the Reynolds number is
evaluated using the height of the step). The velocity profile at
the entrance section represented the Poiseuille's parabola
(\ref{Puaz}). The dimensionless liquid mass flow rate $J$ was
taken equal to 1; it corresponded to the choice of the entrance
pressure gradient in form of
$$ \frac{\partial p}{\partial x}= -
\frac{96}{Re}\Bigl[1+\frac{48\tau}{Re}\Bigr]^{-1}. $$
For the small values of $\tau$ and the big values  of $Re$ we may suppose
that
$$ \frac{\partial p}{\partial x}= - \frac{96}{Re}. $$ The computed length of
the
separation zone behind the step was compared with data from \cite{Sparrow}.
It
was also defined from graphs presented in \cite{Armaly}.

In \cite{Sparrow} the Reynolds number was derived from the average
flow velocity and from the height of the step. The entrance
profile was also set in form of the Poiseuille's parabola. The
 mass flow rate  $J$ was taken equal to 1. The
results, in particular, contain the length of the separation zone
for $H=2h$, $Re(h)$=100, 200, 300. In \cite{Armaly} the Reynolds
number was derived from the value of $2h$ and the average entrance
velocity. Graphic data concerning the length of the separation
zone for $50<Re(2h)<800$ are presented here.

The results, obtained by authors, are systematized in the table
\ref{tabl1}. Here $L$ is the dimensionless length of the
computational domain, $N_x$, $N_y$ are the numbers of mesh points
in both directions, $L_s$ is the length of the separation zone,
$N_{iter}$ is the number of time steps till the conversion is
achieved. The spatial mesh is uniform in both directions with
equal widths $h_x=h_y=0.025$. It is well known that the usage of
equal widths $h_x$ and $h_y$ improves the accuracy of description
of the separating flow.

We have $Re_s \sim 10^6$ in the described flows, so the value $\tau =
\tau_0$ in (\ref{tau1})was taken equal to $ \tau_0 =0.5/Re$. The time step
$\delta t$ was equal to $10^{-4}$ for all variants of computation.

\begin{table}[ht!]
\medskip
\begin{center}
\begin{tabular}{ccccc}
\hline $Re(h)$ & 100 & 200 & 300 & 400 \\ \hline
 $L$ & 7.5 & 5.0 & 7.5 & 10 \\
\hline $N_x \times N_y$ & $300\times40$ & $200\times40$ &
$300\times 40$ & $400\times 40$ \\
 \hline
  $\tau$ & 0.005 & 0.0025 & 0.00166 & 0.00125 \\
\hline
 $N_{iter}$ & $19 800$ &
$\sim 20 000$ & $\sim 60 000$ & $\sim 110 000$\\
 \hline
 $L_s/h$ present comp & 5.0 & 8.2 & 10.1 & 14.8 \\
  \hline
   $L_s/h$ \cite{Sparrow}, comput & 4.43 & 7.5 & 10.0 & - \\
 \hline
  $L_s/h$  \cite{Armaly} exp & 5.0 & 8.5 & 11.3 & 14.2 \\
\hline
  $L_s/h$  \cite{Armaly} comput & 5.0 & 8.3 & 8.4 & 7.8 \\
   \hline\hline
\end{tabular}
\end{center}\caption{Computations of laminar flows}
\label{tabl1}
\end{table}

 Computation stops when the condition $\delta p<10^{-3}$ is satisfied.
 $$
\delta p=\max\Big|\frac{p^{n+1}-p^n}{\delta t}\Big|, $$ $n$ is the time step
number.

In all variants the flow reaches the stationary regime. The length
of the separation zone $L_s$ was defined by the location of the
zero stream-function line. It is indicated with the accuracy 0.2.
Comparison of the results mentioned above with corresponding data
from the Navier-Stokes simulation and with the experiments
\cite{Armaly} demonstrates good agreement both in the length of
the separation zone and in the picture of the flow in general.
Mention, the good agreement for QHD and experimental results for
Re = 400. An almost linear increase of the values of $L_s$ is
observed in computations with the increasing number $Re$.

\begin{figure}[htb!]
\includegraphics[width=.95\textwidth]{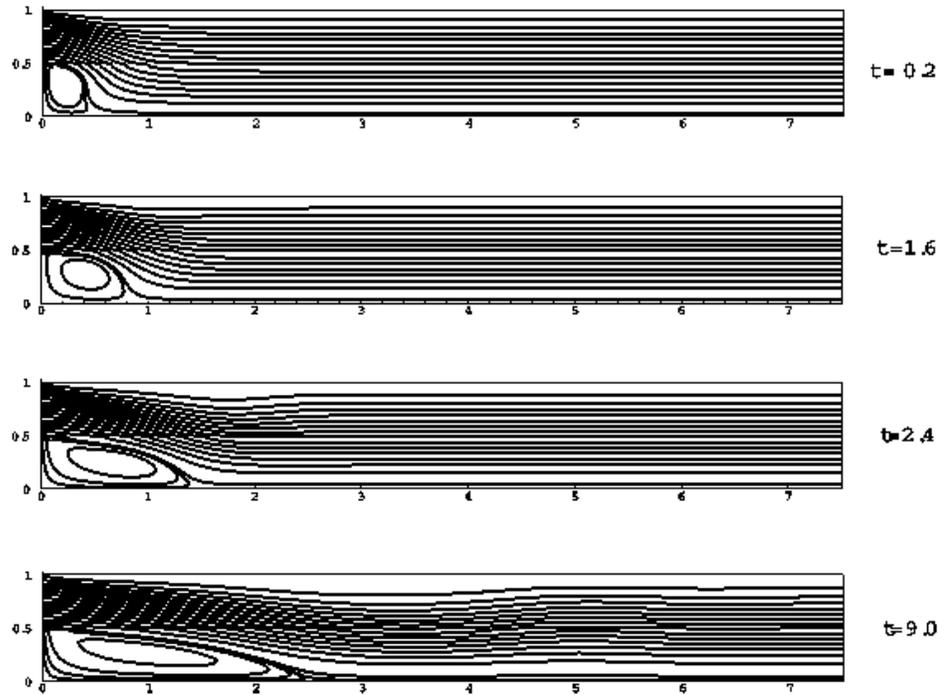}
\caption{Stream functions for Re=100} \label{Fig2}
\end{figure}

\begin{figure}[htb!]
\includegraphics[width=.95\textwidth]{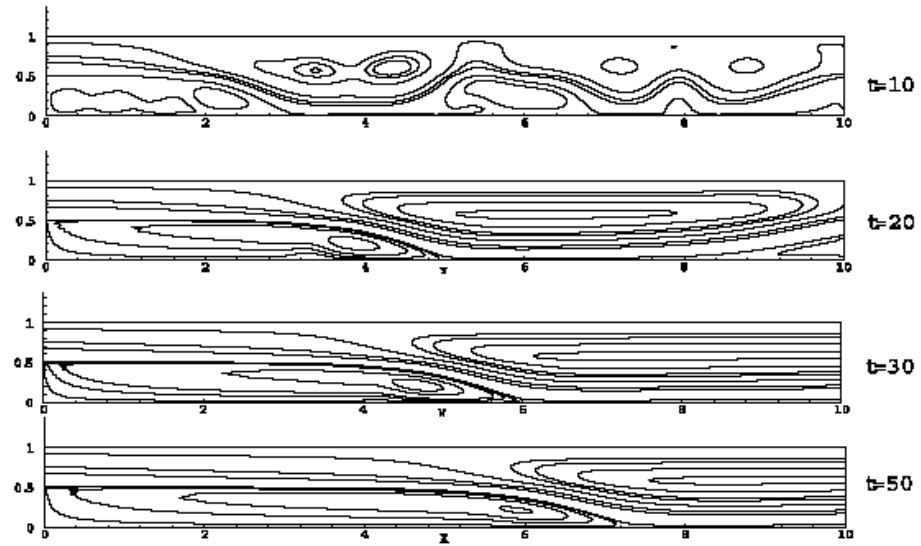}
\caption{Stream functions for Re=400} \label{Fig3}
\end{figure}

For Re = 100 and 200 the process of flow relaxation consists of
the appearence and further growth of a single vortex behind the
step. For Re = 300 and 400 this process proves to be oscillatory
and is accompanied with arising and separation of vortex-like
formations, but, unlike the regimes with greater Reynolds numbers
(they are considered in the second part of this paper), this
oscillations fade and finally form a single stationary vortex
behind the step. The isolines of the flow function $\psi$,
constructed according to (\ref{streamf}), are demonstrated in
Figs. \ref{Fig2}, \ref{Fig3}. They illustrate the process of flow
relaxation in time for $Re=100$ and $400$. The isolines are placed
equidistantly.

With further increasing of the Reynolds number the stationary
solution becomes unstable.

The influence of the regularization parameter $\tau$ and the conversion of
the
numerical solution was investigated for the variant with $Re=100$. The value
of
$\tau$ was additionally chosen equal to $5\cdot 10^{-4}$ and $5\cdot 10^{-
2}$;
the time step $\delta t$ was changed proportionally.

Besides the mesh described in the table, we also used another one
- with twice as many nodes in both directions. The decreasing of
the spatial mesh size  by factor of two caused the analogous
decreasing of the time step. It was shown that the length of the
separation zone and the general picture of the flow practically
doesn't depend neither of the value of regularization parameter
$\tau$ nor of the spatial mesh widths $h_x$ and $h_y$. The
increasing of $\tau$ causes smoothing of the flow picture and
allows us to increase the time step. Spatial mesh refinement gives
a more detailed picture of the flow.

We have studied the dependence of the solution on the pressure
gradient at the entrance section with the average velocity and
mass flow rate remaining constant. It was found out that the
pressure gradient variation in the range from $-96/Re$ to $-12/Re$
practically doesn't influence the structure of the flow: at the
distance around $0.5 h$ from the entrance boundary  the pressure
adjusts to the existing liquid mass flow rate and practically
doesn't depend on the initial gradient.

\section{Conclusion}

The present paper contains the phenomenological derivation of
quasi-hydrodynamic equations. Two-dimensional mathematical model describing
the viscous incompressible flow behind the back step is formulated and solved
numerically.

The computer simulation shows that the flows with small Reynolds numbers that
correspond to the laminar regime, are stationary. The obtained regimes are in
good agreement with the corresponding solutions of the Navier-Stokes system
and
with experimental data mentioned in literature. Oscillations that appear in
the
solutions describing the relaxation of laminar flows for moderate Reynolds
numbers, fade with time. The final flow doesn't depend on the choice of the
smoothing parameter $\tau$, which plays the role of regularizator in these
computations.

These results are in consistence with theoretical estimates
\cite{Sher2000}. According to them, additional QHD-terms should be
small in case of stationary flows and the solution of the QHD
system is expected to be close to the solution of the
Navier-Stokes system. Additional terms act as the regularizators
and allow us to apply a relatively simple, stable and accurate
numerical algorithm.

The authors acknowledge Laboratoire D'Aerothermique du CNRS,
Orleans, and personally Dr. J.-C. Lengrand, for the permanent
support of this research.


\end{document}